\documentclass[final,5p,times,twocolumn]{elsarticle}

 \usepackage{graphics}
 \usepackage{array}

\usepackage{amsmath}
 \usepackage{amsthm}

 \usepackage{lineno}

 \biboptions{comma,square,compress}

\journal{Physics Letters B}

\begin{document}

\title{Quantifying angular distributions in multinucleon transfer reactions with a semi-classical method}
\author[1]{Zehong Liao}
\author[1]{Zepeng Gao}
\author[1]{Yu Yang}
\author[1]{Yueping Fang}
\author[1,2]{Jun Su}
\author[1,2]{Long Zhu\corref{mycorrespondingauthor}}
\cortext[mycorrespondingauthor]{Corresponding author: zhulong@mail.sysu.edu.cn}

\affiliation[1]{organization={Sino-French Institute of Nuclear Engineering and Technology, Sun Yat-sen University},
            city={Zhuhai},
            postcode={519082}, 
            state={},
            country={China}}
\affiliation[2]{organization={Guangxi Key Laboratory of Nuclear Physics and Nuclear Technology, Guangxi Normal University},
            city={Guilin},
            postcode={541004}, 
            country={China}}

\begin{abstract}
The multinucleon transfer (MNT) process in low-energy heavy ion collisions can be utilized to produce unknown nuclei far beyond the stability line. However, the reaction products exhibit broad angular and energy distributions, which could lower the experimental detection efficiency. We present a classical approach that employs a parameterized angular distribution to describe the complex issue. By analyzing limited experimental data on angular distribution, we proposed a three-parameter formula to calculate the angular distribution and identified the dependencies of the parameters. We also discuss the sensitivity of these parameters within this method. A comprehensive comparison with microscopic models and experimental data across a wide range of conditions is conducted. The proposed formula offers an efficient and straightforward way to determine the angular distribution of MNT products.
\end{abstract}

\maketitle


In the 1970s and 1980s, multinucleon transfer (MNT) reactions between massive nuclei began to attract significant attention \cite{GRIDNEV1970385, GALIN1970461, volkov1978DIC}. Since then, the significant experimental and theoretical efforts have been made to understand the characteristics and mechanisms \cite{NORENBERG1974289, PhysRevLett.33.1105, PhysRevLett.32.738, PhysRevC.13.2347}. The complex rearrangement of nucleons that occurs during the evolution of the neck between the colliding partners can be used to produce new isotopes far beyond the stability line \cite{watanabe2015pathway}. The advantages of MNT reactions for synthesizing neutron-rich nuclei have been widely discussed \cite{WOS:000259377500026, bao2018influence, zhu2017theoretical, PhysRevC.107.014614}. However, separating the MNT products with wide angular and energy distributions remains a challenging task, because the reaction products in the MNT reactions exhibit strongly anisotropic angular distributions \cite{PhysRevLett.32.738}.

Accurately describing the differential cross-sections of multinucleon transfer reactions, such as angular distributions, is a theoretical difficulty. Recent years have seen significant theoretical advancements \cite{wen2020dydns, PhysRevC.96.024618, 2020saikomnt, PhysRevResearch.5.L022021, ZHU2024138423, zhang2024multinucleon, dfc2024}, driven by increasing experimental data from large acceptance magnetic spectrometers used in heavy-ion reactions. These advancements have improved the accuracy of reproducing differential cross-section data, such as angular and energy distributions. Nevertheless, the validity of these models, based on the assumptions used to simplify calculations, often introduces the uncertainty of the interaction potential parameter and dynamics frictional form factors. The goal of this work is to develop an efficient, semi-classical deflection function to calculate the global angular distribution in MNT reactions.

Due to high nuclear viscosity and significant overlap of nuclear surfaces, two massive nuclei in the MNT reaction can stick together and rotate at a certain angle. When the Coulomb and centrifugal forces between the nuclei exceed the nuclear attraction, the system breaks apart and separates in the direction of the deflection angle. Wolschin \cite{wolschin1978analysis} observed that the deflection angle $\Theta$ is related to the rotation angle $\Delta \theta$ by
\begin{flalign}
    \begin{split}
        \Theta(l_{i}) = \pi - \theta_{i}(l_{i}) - \theta_{f}(l_{i}) - \Delta \theta, 
    \end{split}\label{eq_angle}
\end{flalign}
where $\theta_{i}(l_{i})$ and $\theta_{f}(l_{i})$ are respectively the Coulomb deflection angles in the entrance and exit channels and $l_{i}$ is the entrance-channel relative angular momentum. The above equation can be written as 
\begin{flalign}
    \begin{split}
        \Theta(l_{i}) = \Theta_{C}(l_{i}) - \Theta_{N}(l_{i}),
    \end{split}\label{eq_deflection}
\end{flalign}
where $\Theta_{C}$ is the Rutherford deflection function and $\Theta_{N}$ is the nuclear part of the deflection function. The Coulomb deflection is given by the Rutherford function as
\begin{flalign}
    \begin{split}
        \Theta_{C}(l_{i}) = 2 \mathrm{arctan}\frac{Z_{1}Z_{2}e^{2}}{2E_{\mathrm{c.m.}}b}.
    \end{split}\label{eq_coulomb}
\end{flalign}

The term $\Theta_{N}$ can be parameterized as  
\begin{flalign}
    \begin{split}
        \Theta_{N}(l_{i}) = \beta\Theta_{C}^{\mathrm{gr}}\frac{l_{i}}{l_{\mathrm{gr}}}(\frac{\delta}{\beta})^{\frac{l_{i}}{l_{\mathrm{gr}}}},
    \end{split}\label{eq_nuclear}
\end{flalign}
where $\Theta_{C}^{\mathrm{gr}}$ is the Coulomb grazing angle at the grazing angular momentum; $l_{\mathrm{gr}}$ is the grazing angular momentum, which can be calculated as $0.22R_{\mathrm{int}}[A_{\mathrm{red}}(E_{\mathrm{c.m.}}-V(R_{\mathrm{int}}))]^{1/2}$. The reduced mass of the projectile and target nuclei is denoted as $A_{\mathrm{red}}$, and $V(R_{\mathrm{int}})$ represents the interaction potential at the interaction radius $R_{\mathrm{int}}$. $\beta$ and $\delta$ are two free parameters that depend on the degree of focusing observed in the experimental angular distribution. $\beta$ and $\delta$ can be further constrained by fitting them to the experimental gross angular distribution using the classical approximation given by

\begin{figure}[htbp]
    \includegraphics[width=8cm]{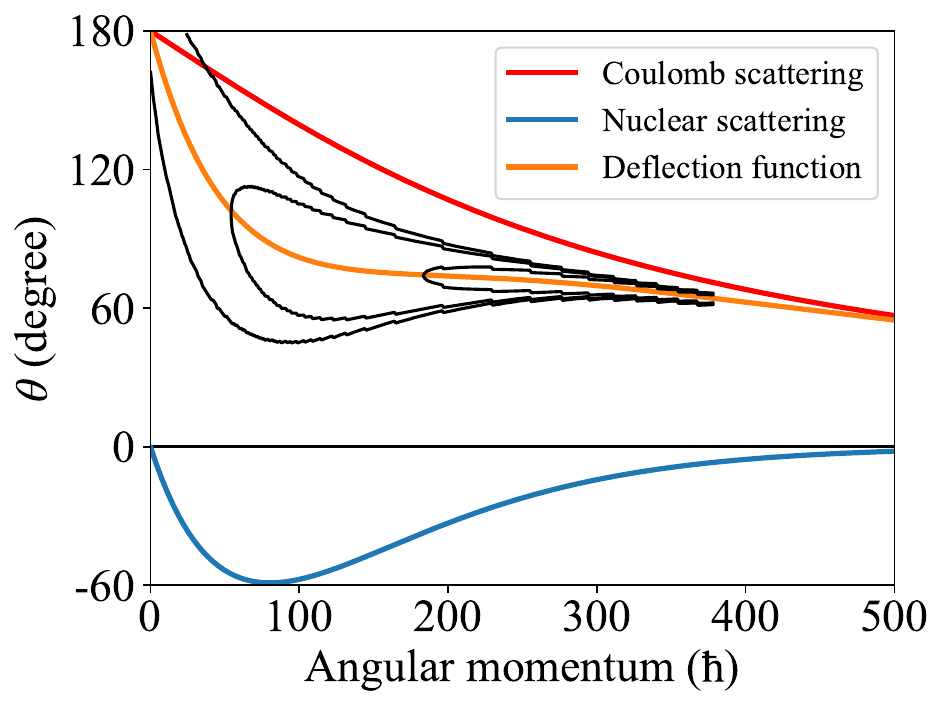}
    \caption{Black contour lines for the probability in the reaction $^{136}\mathrm{Xe}+^{209}\mathrm{Bi}$ at the incident energy $E_{\mathrm{c.m.}}$ = 569 MeV. The orange line corresponds to the mean deflection angle, and the parameters used in this figure are the optimal ones shown below. The Coulomb and nuclear deflection angles as a function of angular momentum are denoted as red and blue lines, respectively.}
    \label{figure1}
\end{figure}

\begin{flalign}
    \begin{split}
        \frac{d \sigma}{d \Theta}=\frac{2\pi }{k^{2}} \sum_n l_n\left|\frac{d l}{d \Theta}\right|_{l=l_n}.
    \end{split}\label{eq_integal}
\end{flalign}

 Li \cite{li1983distribution} investigated the dependence of $\beta$ and $\delta$ on the modified Sommerfeld parameter $\eta^{\prime}$ by comparing angular distributions of various heavy systems to obtain the approximations. The modified Sommerfeld parameter is defined as $\eta^{\prime} = \frac{Z_{1}Z_{2}e^{2}}{v^{\prime}}$, where $v^{\prime}$ represents the velocity at the interaction barrier $V(R_{\mathrm{int}})$. Furthermore, the empirical formula also can be extended to evaluate the interaction time as a function of the initial angular momentum \cite{wolschin1978analysis, W.Li_2003}. Due to the limited availability of experimental data at that time, Li utilized only a few experimental systems and did not mention the limitations of the theoretical calculations, which could affect the fitting results \cite{li1983distribution}. Furthermore, although the peak values matched, the overall fitting of the angular distribution was too narrow compared to the experimental data \cite{wolschin1978analysis}, which may affect the reliability of the deflection function.

\begin{figure*}[htbp]
    \centering
\includegraphics[width=17cm]{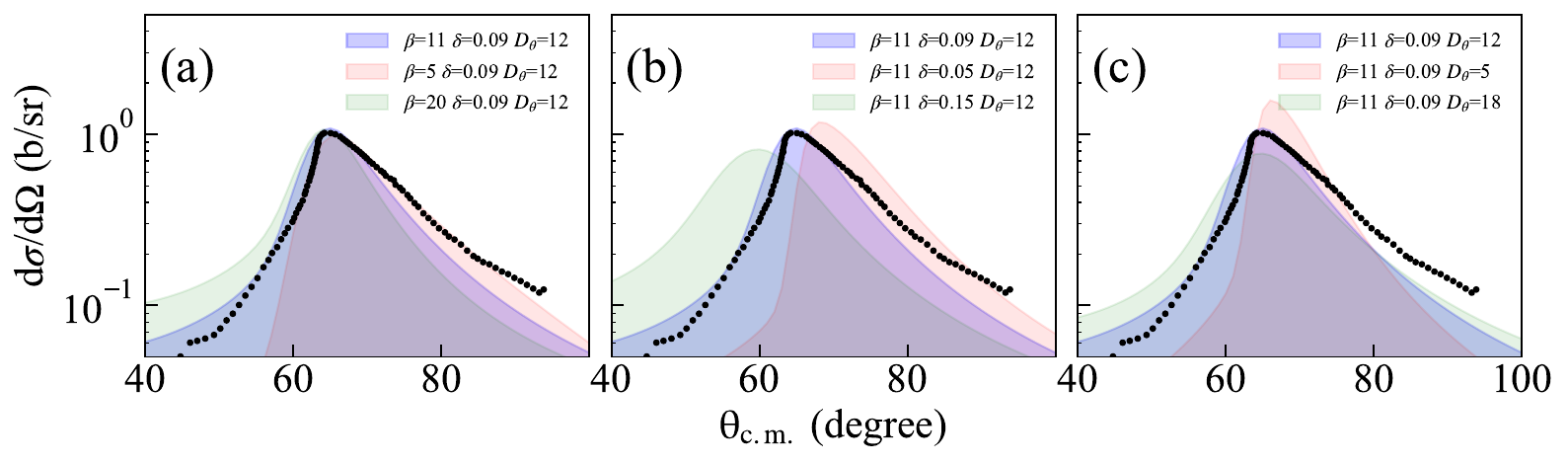}
    \caption{The effects of the parameters $\beta$, $\delta$, and $D_{\theta}$ on the angular distribution in the reaction $^{136}$Xe + $^{209}$Bi at the incident energy $E_{\mathrm{c.m.}} = 569$ MeV. The black circles are taken from Refs. \cite{xebi526684}.
    }
\label{figure2}
\end{figure*}

As we discussed, using the two-parameter deflection function allows us to relate the entrance-channel angular momentum with the exit-channel angle $\theta$. Additionally, fluctuations are highly noticeable \cite{riedel1980statistical, PhysRevLett.105.192701, PhysRevC.79.054606, PhysRevLett.127.222501, 1980Statistical}, especially in the case of violent collisions. Experimentally, the differential cross section \(\frac{\partial^{2} \sigma}{\partial \theta \partial E}\), as depicted in Wilczynski diagrams, sometimes reveals ridges that suggest the validation range of a deterministic trajectory predicted theoretically. This is especially true for small energy losses and for masses that are close to those of the entrance channel. However, fluctuations significantly obscure any clear indication of a deterministic trajectory. This notable evolution is associated with a progressive increase in both interaction time and energy dissipation. In this work, we consider performing a Gaussian expansion of the deflection trajectory provided by this deflection function and adjusting the width of the angular distribution by introducing an additional parameter to control the variance of the Gaussian distribution.

\begin{flalign}
    \begin{split}
        G(\theta)=\frac{1}{\sigma_{\theta}(l_{i})\sqrt{2\pi}}\exp{(-\frac{(\theta-\Theta(l_{i}))^{2}}{2\sigma_{\theta}^{2}(l_{i})})},
    \end{split}\label{eq_gaussi}
\end{flalign}

where the width coefficient $\sigma_{\theta}^{2}(l_{i})$ is a function of the initial relative angular momentum. The value of this width coefficient will be discussed later. The central value $\Theta(l_{i})$ is given by Eq. (\ref{eq_deflection}). 

As illustrated in Fig. \ref{figure1}, we present the deflection function of the reaction $^{136}\mathrm{Xe} + ^{209}\mathrm{Bi}$ at the incident energy $E_{\mathrm{c.m.}} = 569$ MeV. The contributions of Coulomb and nuclear deflection in the scattering process are also depicted. There is only one exit angle for a particular angular momentum, which corresponds to the orange line in Fig. \ref{figure1}. When we apply a Gaussian expansion to this deflection function, indicated by the black contour, the angular distribution broadens as the angular momentum decreases. Statistically, the width around the average scattering angle correlates with the reaction time. This relationship adheres to Einstein's fluctuation-dissipation theorem: the more intense the reaction dissipation and the longer the reaction time, the greater the fluctuation. In this analysis, a linear dependence of the interaction time on the scattering angle has been assumed \cite{PhysRevC.16.623, PhysRevC.11.1265} :
\begin{flalign}
    \begin{split}
        \tau_{\mathrm{int}}(l_{i}) = \frac{I_{\mathrm{tot}}}{l_{i}\hbar}(\Theta_{C}(l_{i}) - \Theta(l_{i})),
    \end{split}\label{eq_time}
\end{flalign}
where $I_{\mathrm{tot}}$ is the moment of inertia of the double nucleus system. According to the Fokker-Planck equation with constant transport coefficients, the variance of the angular distribution increases linearly with time. This variance, as a function of the initial angular momentum, can be written as:
\begin{flalign}
    \begin{split}
        \sigma_{\theta}^{2}(l_{i}) = 2 D_{\theta} \tau_{\mathrm{int}}(l_{i}).
    \end{split}\label{eq_sigma}
\end{flalign}
Here $D_{\theta}$ is the diffusion coefficient, which is adjusted to fit the angular distribution. The equations described above allow us to determine the global angular distribution by adjusting three parameters: $\beta$, $\delta$, and $D_{\theta}$. For the reaction $^{136}$Xe + $^{209}$Bi ($E_{\mathrm{c.m.}} = 569$ MeV), we found that the parameters $\beta = 11$, $\delta = 0.09$, and $D_{\theta} = 12$ most closely matched the experimental data. Generally, the parameter $\beta$ determines the depth of the deflection function, while $\delta$ determines the deviation of the scattering angle from the Coulomb grazing angle. For the heavy system $^{136}\mathrm{Xe} + ^{209}\mathrm{Bi}$ we studied, strong focusing can be observed in the deflection function, which is consistent with the experimental angular distribution characteristics of heavy systems.

In Fig. \ref{figure2}, we apply this procedure and different sets of parameters to calculate the angular distribution in the reaction $^{136}$Xe + $^{209}$Bi ($E_{\mathrm{c.m.}} = 569$ MeV), where the experimental data \cite{xebi526684} provide complete and sufficiently precise information on the gross angular distribution. Due to the total kinetic energy loss (TKEL) truncation of the experimental data, for example in this experiment $^{136}$Xe + $^{209}$Bi, the detected data is in this range 309 MeV $>$ TKEL $>$ 23 MeV. Therefore, some cut-off conditions need to be considered in the calculations of the angular distributions. The upper limit of our angular momentum integral is roughly estimated by this condition
\begin{flalign}
    \begin{split}
        \sigma_{\mathrm{exp}} = \frac{2\pi }{k^{2}}\sum_{n}^{l_{\mathrm{cut}}} l_n,
    \end{split}\label{eq_sigma_expand}
\end{flalign}
where $l_\mathrm{cut}$ is the upper limit of the collision relative angular momentum in the process of fitting the angular distribution, ensuring that the fitted cross-section is consistent with the experimental cross-section. This approximation is also used for parametric fitting of other systems in this work.

Understanding the effects of these parameters on the angular distribution is the first step. Figs. \ref{figure2}(a), (b), and (c) show the effects of the parameters $\beta$, $\delta$, and $D_{\theta}$ on the angular distribution, respectively. In Fig. \ref{figure2}(a), we chose two other different values $\beta = 5$ and $\beta = 20$ to compare with the optimal value of the parameter $\beta = 11$ while keeping other parameters unchanged. It is clear that the parameter $\beta$ does not significantly affect the peak value of the angular distribution but has a great impact on the distribution in the front angle region. Larger values of $\beta$ lead to more forward angle scattering, while smaller values of $\beta$ result in smaller forward angle contributions, where the nuclear rainbow of the deflection function moves close to or coincides with the Coulomb rainbow.

For the parameter $\delta$ shown in Fig. \ref{figure2}(b), 
it is evident that $\delta$ is associated with the peak of the angular distribution. This is mainly because the deviation from the Coulomb trajectory near the grazing angle is determined by $\delta$. Larger values of $\delta$ lead to a more left-sided peak of the angular distribution compared to the experimental data. In addition, the physical significance of the parameter $D_{\theta}$ becomes more apparent, as it controls the width of the average scattering angle under the deflection function governed by the first two parameters $\beta$ and $\delta$. As shown in Fig. \ref{figure2}(c), the higher the value of the parameter $D_{\theta}$, the wider the angular distribution, and vice versa.

\begin{figure}[htbp]
    \centering
    \includegraphics[width=8cm]{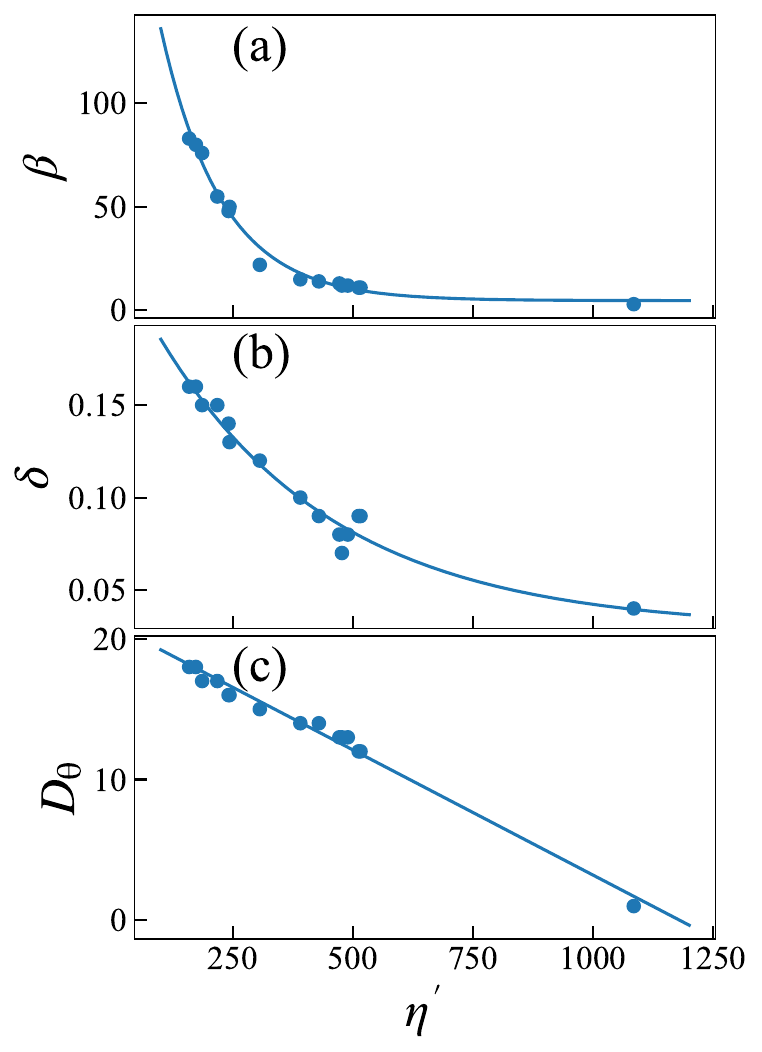}
    \caption{(a-b) Parameters $\beta$ and $\delta$ determine the mean deflection angle $\Theta(l_{i})$ as functions of $\eta^{\prime}$. (c) Parameter $D_{\theta}$ determines the fluctuation around the mean deflection angle as a function of $\eta^{\prime}$. Blue symbols represent fits to the angular distributions, while the solid curves represent the analytical expressions given in the text.}
    \label{figure3}
\end{figure}

Given the scarcity of MNT experiments and the limited availability of angular distribution data, only 17 sets of experimental data were utilized in this study. Employing the method outlined earlier, we adjusted the three parameters to align with the overall experimental angular distribution, resulting in a total of 17 parameter sets. It was observed that $\beta$, $\delta$, and $D_{\theta}$ exhibit smooth variations as functions of $\eta^{\prime}$, as illustrated in Fig. \ref{figure3}. In the fitting process, we found that $\delta$ and $D_{\theta}$ related to each other. This relation primarily arises from the parameter $\delta$ also controls the width of the angular distribution to a certain extent. By fitting the dependence of $\beta$, $\delta$, and $D_{\theta}$ on $\eta^{\prime}$, we obtain the following approximations

\begin{flalign}
&\left\{
\begin{aligned}
&\beta(\eta^{\prime}) = 4.7 + 81.2 \cdot \exp\left[-\frac{\eta^{\prime} - 160.8}{126.5}\right]; \\
&\delta(\eta^{\prime}) = 0.029 + 0.127 \cdot \exp\left[-\frac{\eta^{\prime} - 176.3}{364}\right]; \\
&D_{\theta}(\eta^{\prime}) = 21 - 0.0178\eta^{\prime}.
\end{aligned}
\right. &
\end{flalign}

From Fig. \ref{figure3}, it can be seen that there are minor discrepancies between the data points and the fitted function. Our objective is to establish an empirical parametric formula that accurately describes the overall angular distribution pattern, and these deviations fall within acceptable limits for this study.

\begin{figure}[htbp]
    \centering
    \includegraphics[width=8cm]{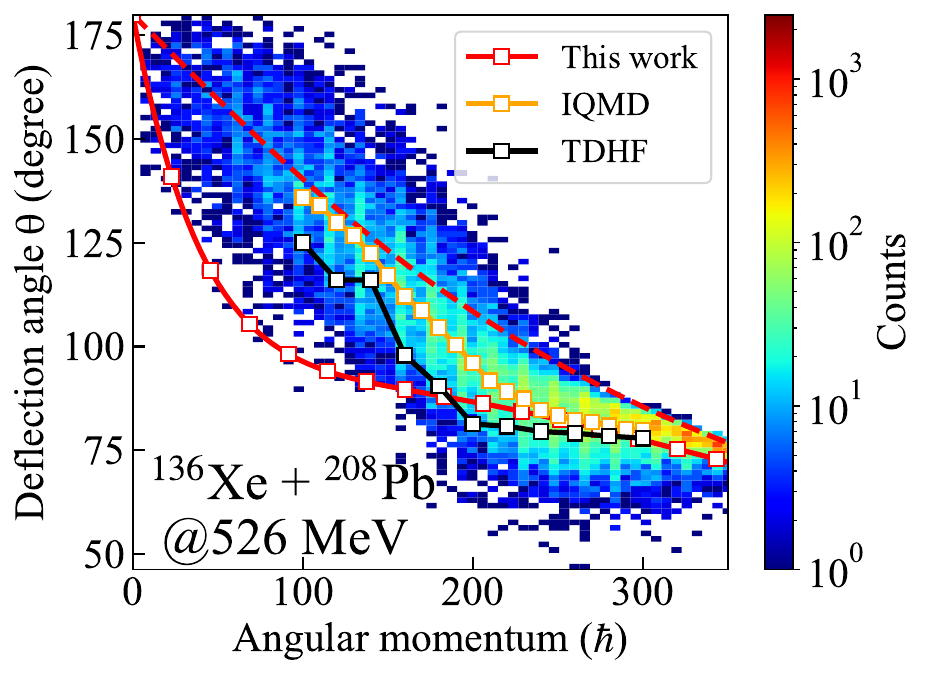}
    \caption{
    Model comparison of the deflection angle as a function of the relative angular momentum for the reaction $^{136}\mathrm{Xe}$+$^{208}\mathrm{Pb}$ at the incident energy $E_{\mathrm{c.m.}}$ = 526 MeV. The theoretical result was obtained using the TDHF (Black open squares), IQMD (Yellow open squares), and newly developed formula (Red open squares).
    }
    \label{figure4}
\end{figure}

To illustrate the effectiveness of the newly developed semi-classical method in this study, we conducted a comparative analysis using the $^{136}\mathrm{Xe}$+$^{208}\mathrm{Pb}$ reaction at $E_{\mathrm{c.m.}}$ = 526 MeV. The primary fragment results from time-dependent Hartree-Fock (TDHF) theory \cite{PhysRevC.109.L041605} and the isospin-dependent quantum molecular dynamics (IQMD) model \cite{PhysRevC.60.064604} are employed. The comparison of deflection angles as a function of angular momentum is depicted in Fig. \ref{figure4}. The TDHF model provides a deterministic trajectory of the deflection angle. The mean value of the deflection angle from the IQMD model is displayed by yellow squares. Based on the numerous simulated results of the IQMD model, the contour plot shows the count of the results on a logarithmic scale. As the angular momentum decreases, it is evident that the calculated deflection angles from all models gradually deviate from the Coulomb deflection angle indicated by the red dashed line. However, upon approaching central collisions, these angles gradually converge back towards the Coulomb deflection angle and eventually reach 180 degrees. This trend reflects the changes in the dominant role of nuclear and Coulomb forces at different distances. The comparison reveals that the calculations from all three methods yield relatively similar results when the relative angular momentum $L>$ 200 $\hbar$. Actually, for depicting the general shape of the angular distribution, the contributions from large angular momentum dominate. 

\begin{figure*}[htbp]
    \centering
    \includegraphics[width=17cm]{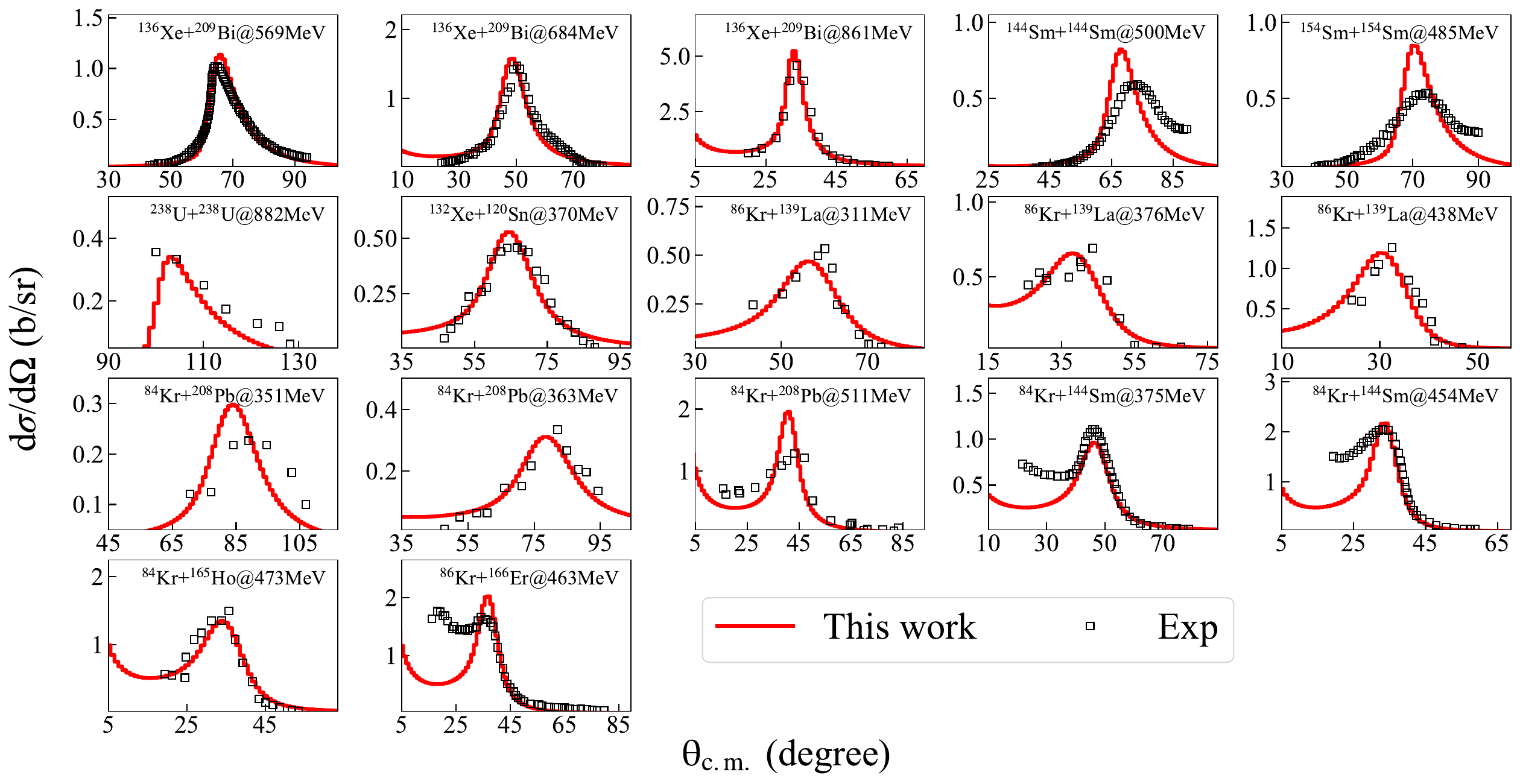}
    \caption{
    The experimental and predicted angular distribution. The results obtained from the parametrized angular distribution formula are shown with the red line, while the black circles denote experimental data taken from Refs.\cite{xebi526684, xebi861, PhysRevC.29.158, BECK197835, PhysRevC.14.143, PhysRevC.17.1672,RUDOLF1981109, Schmidt1978MassTA, HILDENBRAND1983179}.
    }
    \label{figure5}
\end{figure*}

\begin{figure}[htbp]
    \centering
    \includegraphics[width=8.5cm]{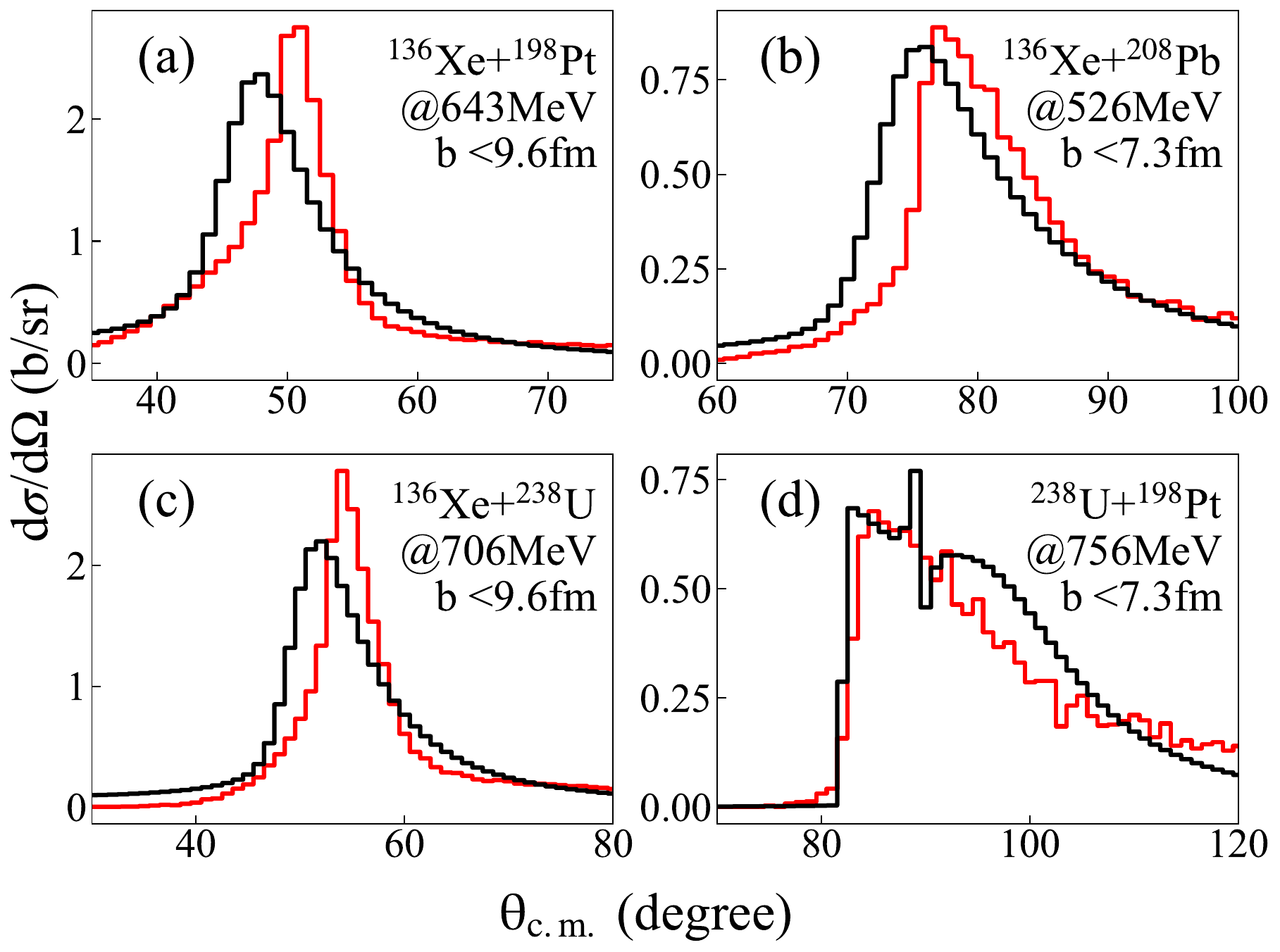}
    \caption{
        Model comparison of the angular distributions for the reactions (a) $^{136}\mathrm{Xe}$+$^{198}\mathrm{Pt}$, (b) $^{136}\mathrm{Xe}$+$^{208}\mathrm{Pb}$, (c) $^{136}\mathrm{Xe}$+$^{238}\mathrm{U}$, and (d) $^{238}\mathrm{U}$+$^{198}\mathrm{Pt}$. The results from our semi-classical approach are represented by black lines, while those from the IQMD model are shown in red lines.
    }
    \label{figure6}
\end{figure}

To validate the effectiveness of the parameterized formula in describing angular distributions, we compared its calculated results with experimental data shown in Fig. \ref{figure5}. The quasi-classical method demonstrates reliable predictive capability, showing good agreement between predicted angular distribution values and experimental data across most reaction systems. However, discrepancies are observed in the forward-angle regions of some light projectile-target combinations, primarily due to the presence of a significant nuclear rainbow effect in such systems. Minor adjustments may be necessary as additional reactions are analyzed. For sufficiently heavy systems, where the deflection function lacks a nuclear rainbow angle, the Coulomb force predominates. It is clearly shown that our method effectively describes angular distributions on both sides of the peak value.

Given the challenges in efficiently separating and detecting neutron-rich unknown isotopes in experiments, the overall fragment angle distribution predicted by the model provides crucial information for optimizing detector placement in MNT experiments. In previous analyses, the capability of the parameterized angular distribution was demonstrated through comparison with existing experimental data. For systems lacking experimental angular distribution data from MNT, the IQMD model was employed to validate the extrapolation capability of the quasi-classical method. For several prominent MNT reaction systems $^{136}\mathrm{Xe}$+$^{198}\mathrm{Pt}$ at $E_{\mathrm{c.m.}}=$ 643 MeV, $^{136}\mathrm{Xe}$+$^{208}\mathrm{Pb}$ at $E_{\mathrm{c.m.}}=$ 526 MeV, $^{136}\mathrm{Xe}$+$^{238}\mathrm{U}$ at $E_{\mathrm{c.m.}}=$ 706 MeV, and $^{238}\mathrm{U}$+$^{198}\mathrm{Pt}$ at $E_{\mathrm{c.m.}}=$ 756 MeV \cite{xebi526684, PhysRevC.86.044611, watanabe2015pathway, PhysRevLett.130.132502}, we compare the angular distributions obtained using the IQMD model with those from our parameterized formula in Fig. \ref{figure6}. To maintain consistency in the calculated cross-section values between the two models, we set the upper limit of the collision parameters as $b < b_{\mathrm{grazing}}$. The comparison reveals that the results obtained in this study, particularly in terms of width, closely resemble those from the IQMD model, albeit with slight deviations in peak position.

In summary, we have introduced and validated a method for studying angular distributions in multinucleon transfer (MNT) reactions. Utilizing the limited experimental data available for MNT angular distributions, we adjusted the key parameters $\beta$, $\delta$, and $D_{\theta}$ to construct the deflection function. This approach circumvents the need for complex dynamic calculations that could introduce additional uncertainties related to potential parameters and frictional form factors.
Our method shows reasonable and consistent results in comparison with the two microscopic approaches TDHF and IQMD in a representative reaction $^{136}\mathrm{Xe}$+$^{208}\mathrm{Pb}$ at $E_{\mathrm{c.m.}}=$ 526 MeV. The three approaches generally produced consistent results for large angular momentum, but significant deviations were observed at small angular momentum. Besides, a systematic comparison involving a diverse dataset encompassing various bombarding energies and reaction systems would provide further clarity on the effectiveness of our method. The robust performance of the method in predicting angular distributions offers a new avenue for investigating other reactions of interest. To demonstrate the extrapolation capability of the method, we employed our model and the IQMD model to analyze the gross angular distributions of reactions such as $^{136}\mathrm{Xe}$+$^{198}\mathrm{Pt}$, $^{136}\mathrm{Xe}$+$^{208}\mathrm{Pb}$, $^{136}\mathrm{Xe}$+$^{238}\mathrm{U}$, and $^{238}\mathrm{U}$+$^{198}\mathrm{Pt}$. Our model effectively captures the characteristic features of angular distributions in deeply inelastic collisions between heavy nuclei, including the transition from weakly focused, with a contribution from negative angle scattering, to strongly focused at positive angles, with increasing system size or decreasing energy. For the currently widely used dinuclear system model, this method can be a more reliable option to be implemented in extended dynamics calculations. 

\section*{Acknowledgements}
The authors would like to thank Pei-Wei Wen for helpful discussion and suggestions. This work was supported by the National Natural Science Foundation of China under Grants No. 12075327 and 12335008; The Open Project of Guangxi Key Laboratory of Nuclear Physics and Nuclear Technology under Grant No. NLK2022-01; Fundamental Research Funds for the Central Universities, Sun Yat-sen University under Grant No. 23lgbj003. The work is supported in part by National Key R$\text{\&}$D Program of China (2023YFA1606402).


\end{document}